\documentclass{article}
\usepackage[english]{babel}
\usepackage[utf8]{inputenc}
\usepackage{johd}
\usepackage{eurosym}
\usepackage{booktabs}
\usepackage{amsmath}

\title{Local energy communities optimization considering cost and greenhouse gases minimization}

\author{S. Barja-Martinez$^{a}$$^{*}$, M. Aragüés-Peñalba $^{a}$, A. Sumper$^{a}$, R. Villafafila-Robles$^{a}$\\
\\
        \small $^{a}$Centre d’Innovació Tecnològica en Convertidors Estàtics i Accionaments (CITCEA-UPC) \\
        \small $^{*}$Corresponding author: \tt{sara.barja@upc.edu} \\
}
\date{}

\begin{document}
\maketitle

\begin{abstract} 
Local Energy Communities (LECs) facilitate consumer involvement in local electricity generation and distribution, offering a significant opportunity for society to participate in the energy transition. This paper presents the optimization of a renewable energy community in Spain, consisting of four office buildings, a collectively owned centralized photovoltaic system, and a Li-ion battery storage system. The case study assesses the performance and feasibility of the proposed solutions. The results indicate a 6\% reduction in emissions and a 20\% reduction in electricity costs, demonstrating the potential of LECs to enhance energy security by saving costs and emission while mitigating the vulnerability of local areas to power outages and disruptions.
\end{abstract}

\noindent\\

\section{Introduction}

The generation and use of energy account for more than 75\% of the EU’s greenhouse gas emissions. To achieve the goal of decarbonizing the economy and reaching net-zero emissions by 2050, the European Green Deal \cite{europeanCommissionGD} proposes various objectives related to buildings and the residential sector:

\begin{itemize}
    \item Enhancing the energy efficiency of buildings.
    \item Ensuring a secure and affordable EU energy supply.
    \item Prioritizing energy efficiency, improving the energy performance of our buildings and developing a power sector based largely on renewable sources.
\end{itemize}

These objectives can be achieved through a digitalized energy sector based largely on renewable distributed energy resources.  Local Energy Communities (LECs) encourage the participation of consumers in the electricity generation and distribution at local level, providing an opportunity for society to be involved in the energy transition. 
According to the European Commission, a LEC is an open and voluntary association that combines non-commercial aims with environmental and social community objectives. The aim is to promote community-driven and decentralized electricity generation, rather than central generation managed by a small number of large power plants, as has been the case until now. 

To take advantage of user participation in local energy communities, electrification at the user's premises is essential; for instance, replacing a conventional natural gas boiler with heat pumps for space heating. Additionally, by increasing electrification, consumers and energy communities can offer flexibility to the energy system through demand response and storage programs. 

The Instituto para la Diversificación y Ahorro de la Energía (IDAE) in Spain, outlines the benefits of local energy communities profiling, including: \cite{Comunida88:online}

\begin{itemize}
    \item Providing citizens with fair and easy access to local renewable energy resources and the opportunity to benefit from these investments.
    \item Empowering users to take control and greater responsibility for meeting their energy needs.
    \item Creating investment opportunities for citizens and local businesses. 
    \item Offering communities to generate income, increasing the acceptance of local renewable energy development.
    \item Enabling the integration of renewable energy into the system through demand-side management.
    \item Environmental benefits.
    \item Social benefits. The creation of local employment and promotion of social cohesion and equity. 
\end{itemize}

The increasing growth and interest in LECs is primarily due to rising electricity prices and an increasing awareness of climate change in society, combined with the growing availability of affordable, small-scale distributed energy resources (DERs). Legislative changes and government subsidies have also helped to accelerate the creation of energy communities and self-consumption generation.

It is a fact that prosumers and their collective forms will play a key role in the forthcoming years by empowering consumers, boosting energy efficiency, building interconnected energy systems that allow peer-to-peer energy trading and better-integrated grids to support renewable energy sources. This contributes to a fairer transition to climate neutrality that allows citizens to take ownership of energy consumption and production.

\section{LEC regulation}

The Clean Energy Package (CEP) \cite{EUcleanpackage} introduced by the European Commission has established a legislative framework for the operation of Local Energy Communities across Europe, which aims to facilitate citizens' participation in energy markets, evolving from traditional passive consumers to prosumers. A report made by the European Commission about community renewable energy in Europe confirms this transition, stating that by 2030 energy communities could own 17\% of installed wind capacity and 21\% of solar Europe-wide. By 2050, almost half of EU households are expected to produce clean energy \cite{Caramizaru2020}.

The CEP introduces two types of energy communities:

\begin{itemize}
    \item Citizen Energy Community (CEC). The \textit{Internal Electricity Market Directive (EU) 2019/944} \cite{DIRECTIV12:online} introduces CEC as a legal entity that is based on voluntary and open participation and is effectively controlled by members or shareholders that are natural persons, local authorities, including municipalities, or small enterprises. CEC constitute a new type of entity due to their membership structure, governance requirements and purpose.
    \item Renewable Energy Community (REC). The \textit{Renewable Energy Directive (EU) 2018/2001} \cite{DIRECTIV13:online} defines a REC as a legal entity that, in accordance with the applicable national law, is based on open and voluntary participation, is autonomous, and is effectively controlled by shareholders or members that are located in the proximity of the renewable energy projects that are owned and developed by that legal entity.
\end{itemize}

These two EU directives establish a legal framework for collective citizen participation in the energy system. The definition of CEC is similar to REC, but there are some fundamental distinctions. RECs have a specific focus on renewable sources and should be located close to renewable energy projects, while CEC has no such restriction. Another difference is that an energy community can only be called a REC if its activity is based on renewable energy sources, while a CEC may use renewable or conventional sources \cite{Biresselioglu2021}. For this article, a REC is presented. The main differences between the two types of energy communities are summarized in Table \ref{tab:rec-cec-comparison}.

{
\renewcommand{\arraystretch}{1.2} 
\begin{table}[h]
\centering
\footnotesize
\begin{tabular}{p{1.5cm}p{4cm}p{4cm}} 
\hline
& \textbf{Citizen Energy Community} & \textbf{Renewable Energy Community} \\
\hline
\textbf{Members} & Natural persons, local authorities, small/microenterprises & Same as CEC, but members' main activity must not be defined by REC membership \\
\hline
\textbf{Location} & No location limits, cross-border possible & Members must be close to the REC project \\
\hline
\textbf{Activities} & Energy sector for members; electricity sector for whole market & Renewable energy in all areas of the energy market \\
\hline
\textbf{Technology} & No limitations & Only renewable energy \\
\hline
\end{tabular}
\caption{Comparison of Citizen Energy Community and Renewable Energy Community \cite{CEER2019}.}
\label{tab:rec-cec-comparison}
\end{table}
}

\begin{figure}[htbp]
  \centering
  \includegraphics[width=0.5\columnwidth]{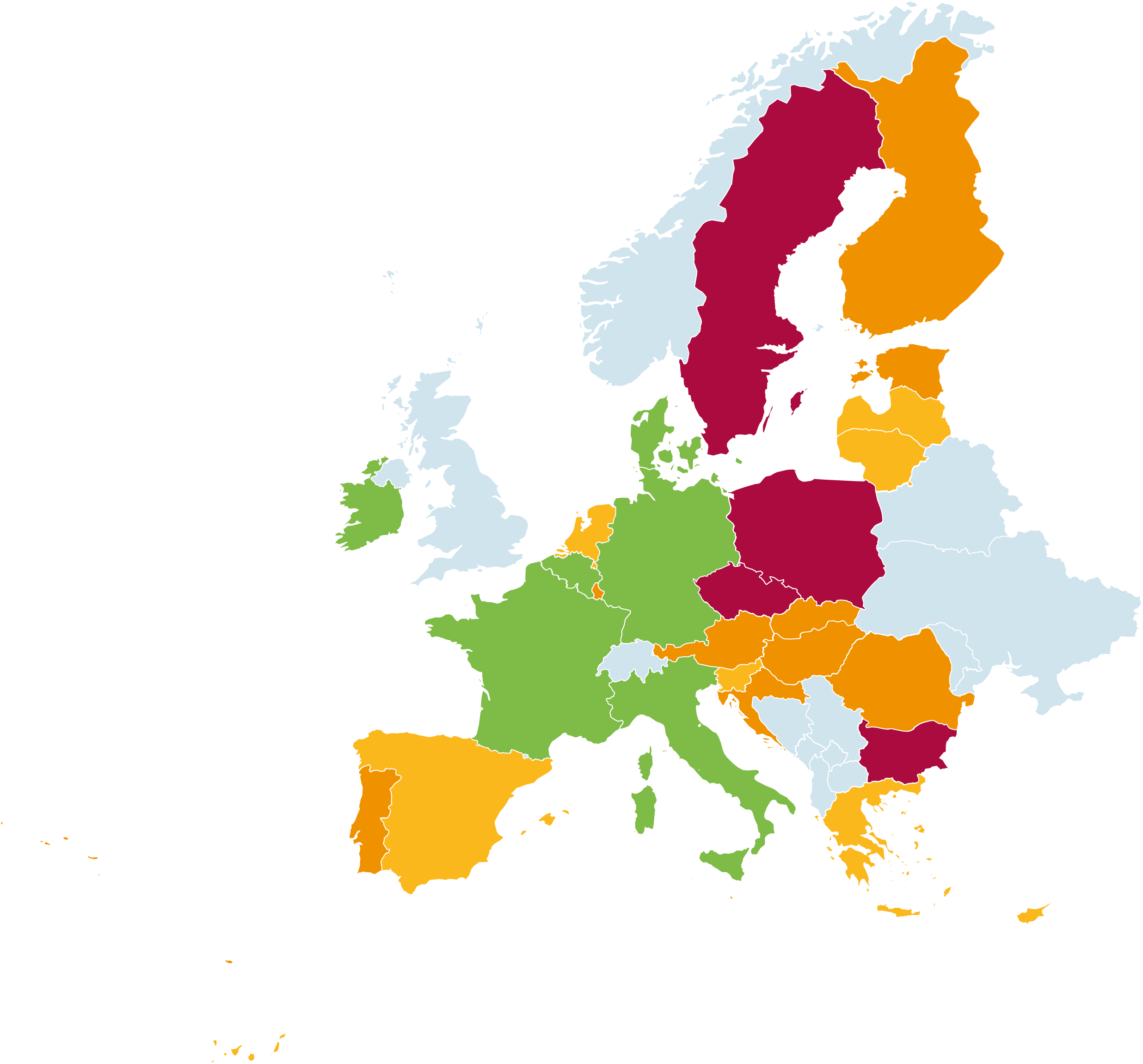}
  \caption{Comparative assessment of the progress for enabling regulation frameworks for RECs in the diverse European countries following a traffic light grading system in which red stands for bad transposition and green for best practices \cite{Guo2022}. Source: REScoop.}
  \label{fig:cecrec}
\end{figure}

However, there are still several challenges confronting the proliferation of LECs. A significant challenge are the regulatory barriers due to the complex framework required to obtain the necessary permits and approvals for installation and operation. This can make it difficult for LECs to secure financing and access to the energy grid, as conventional investors may be uncertain about investing in short-time tested business
models. This is reflected in Figure \ref{fig:cecrec}, which demonstrates that the transposition for enabling frameworks and support schemes for energy communities is not consistent across European member states. This map provides a comparative assessment of this progress, using a traffic light grading system to represent how far each country has progressed towards transposing EU regulations on energy communities.


\subsection{Spanish regulatory framework}


This article focuses on optimizing a renewable energy community located in Spain. To understand the context in which the proposed case study is developed, the fundamental regulatory aspects are explained below.

The Spanish Government has introduced the definition of REC and implemented policies to promote renewable energy development and encourage citizens' participation in the energy system. Most existing renewable energy communities in Spain use the legal framework provided in Real Decreto (RD) 244/2019 \cite{Disposic10:online} for individual and collective electricity self-consumption and the use of renewable energy sources. It defines collective self-consumption as a group of owners sharing one or several solar panel installations. This limits the scope of energy sharing, in particular excluding other renewable technologies such as wind and small-hydro. As shown in Figure \ref{fig:cecrec}, Spain is in "average progress" in implementing the European directives in its national law framework; therefore there is still way to go.

According to RD 244/2019, a REC installation must comply with at least one of the following requirements:

\begin{itemize}
    \item Self-consuming owners must be connected to the same LV transformation center.
    \item The distance between the self-consumer and the energy production center should be no more than 500 meters. The radius was recently expanded to 1km; however, this additional distance is only available to PV self-consumption if located on buildings.
    \item The photovoltaic production facility and the self-consumers must share the same cadastral reference.
\end{itemize}

There are two connection modalities available:

\begin{itemize}
    \item Collective self-consumption with a connection through the public grid: The PV production is shared via the public grid. It is connected to the LV grid through a bi-directional smart meter, and the retailer compensates the end-user(s).
    \item Collective self-consumption with direct connection to the internal grid: In this case, the photovoltaic installation does not connect to the public grid, but the photovoltaic production is distributed directly to each of the internal grids of the self-consumers. This connection is typically used in large industrial customer installations.
\end{itemize}

The following modalities specified in RD 244/2019 require all consumers to belong to the same self-consumption modality. These modalities are: 

\begin{itemize}
    \item Collective self-consumption without surpluses: an anti-spill system is used to avoid injecting surplus energy into the electricity grid.
    \item Collective self-consumption with surpluses not subject to compensation: the owner of the generation facility sells the surplus energy to the electricity market.
    \item Collective self-consumption with surpluses subject to compensation: consumers receive financial compensation for the surpluses they inject into the electricity grid. The retailer is responsible for compensating the surplus energy cost at the end of each billing period. Surpluses that exceed imported consumption are not compensated.
    
\end{itemize}


\textbf{Sharing strategies}

The distribution of PV generation among end-users is determined by the "Sharing Agreement". The RD 244/2019 sets guidelines for distributing the generation among customers, with a fixed distribution coefficient value assigned to each user. An amendment to the regulation permits the establishment of variable distribution coefficients for each user during each hour of the year while still maintaining the option of using fixed coefficients. Once the coefficient is determined for each user, it remains fixed and cannot be changed until one year has passed.



To conclude this section, it is essential to remember and emphasize that regulations for Energy Communities are still evolving and can vary widely between countries and regions in a brief period of time.

\section{Methodology}

\begin{figure}[htbp]
  \centering
  \includegraphics[width=0.75\textwidth]{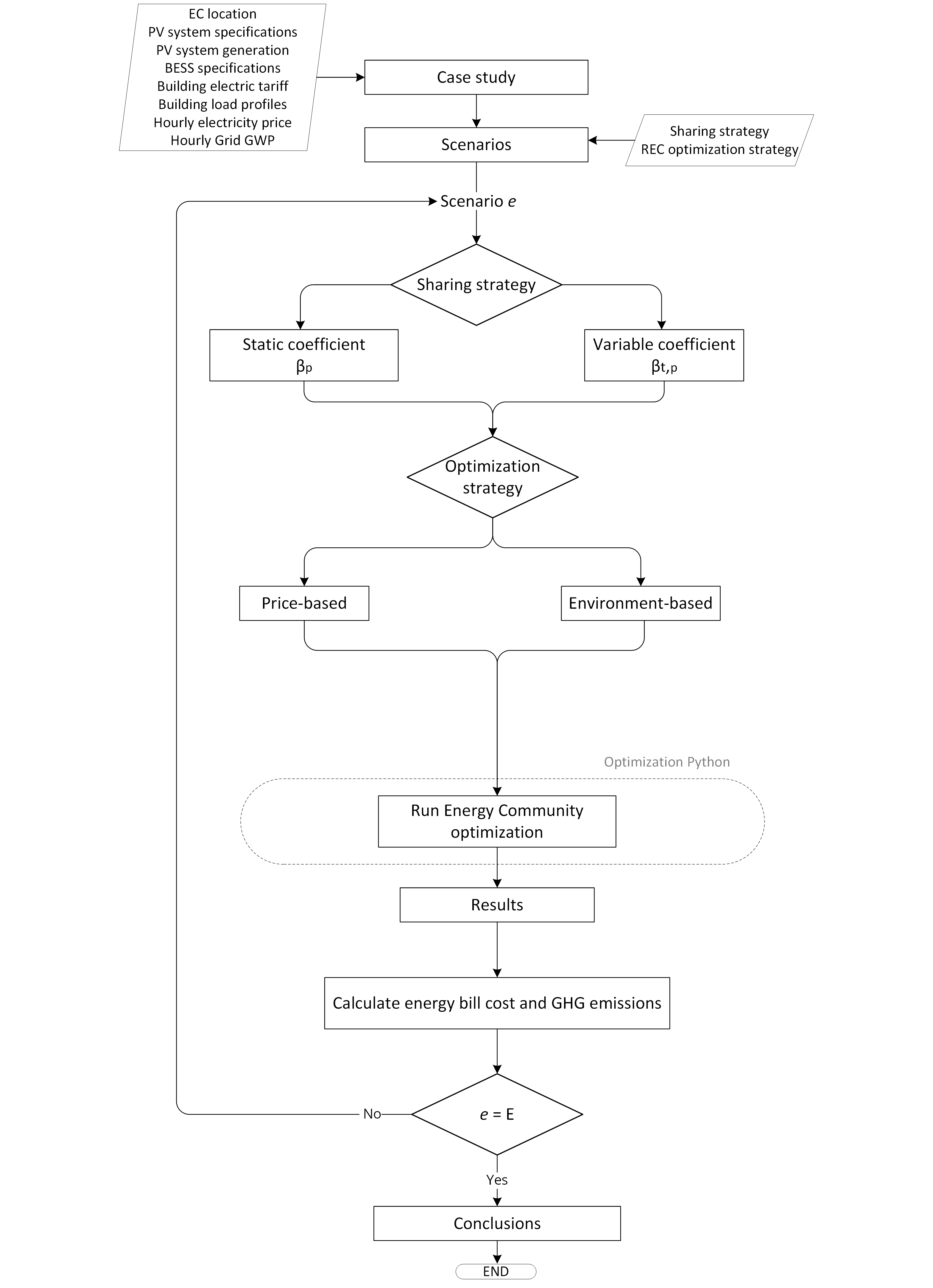}
  \caption{LEC methodology.}
  \label{fig:ECmethodology}
\end{figure}

This section takes into account the existing regulatory framework in Spain and presents the methodology developed to optimise the operation of a Local Energy Communitiy. A scheme of the method followed is presented in Figure \ref{fig:ECmethodology}.

The methodology begins by setting up the case study and scenarios. The required input data includes the location of the renewable energy community, specifications of the BESS and PV systems, the number of electrical supply endpoints and their consumption profiles, PV system generation, hourly electricity price tariff for each end-user and grid GWP. The scenarios are established based on two variables, each linked to a specific analysis: sharing coefficient strategy and the REC optimization strategy. All possible combinations of these variables are established to generate a comprehensive set of scenarios for evaluation and discussion. The results will demonstrate the potential cost savings for the overall REC, as well as the $CO_{2}$ emissions avoided through the optimization strategies designed in this paper.

Regarding the LCA analysis, Table \ref{tab:gwp_lec} displays the range of GWP indicators for the overall life cycle stages for each electricity generation source type, as reported in \cite{Turconi2013}. 

{
\renewcommand{\arraystretch}{1.75}
\begin{table}[htbp]
  \caption{Lyfe cycle emission factors for electricity generation sources \cite{Turconi2013, Thomas2020}.}\label{tab:gwp_lec}
\centering
\footnotesize 
\begin{tabular}{p{2.5cm} p{2.8cm} p{2.8cm}}
\hline
\textbf{Generation source $GS_{i}$} &\textbf{GWP~range [kg~CO$_{2-eq}$/kWh]} & \textbf{Average~GWP [kg~CO$_{2-eq}$/kWh]} \\ \hline
Hard coal  & 0.660-1.05 &   0.855  \\
Lignite  & 0.800-1.30 &    1.05    \\
Natural gas   & 0.38-1 &    0.690    \\
Nuclear    & 0.003-0.035 &    0.019   \\
Biomass    & 0.008-0.130 &        0.069        \\
Hydro-power    & 0.002-0.02   &    0.011   \\
Wind     & 0.003-0.041    &   0.022    \\ 
Battery \cite{Thomas2020}  & - & 0.060\\ \hline
\end{tabular}
\end{table}}

\begin{figure}[htbp]
  \centering
  \includegraphics[width=0.5\textwidth]{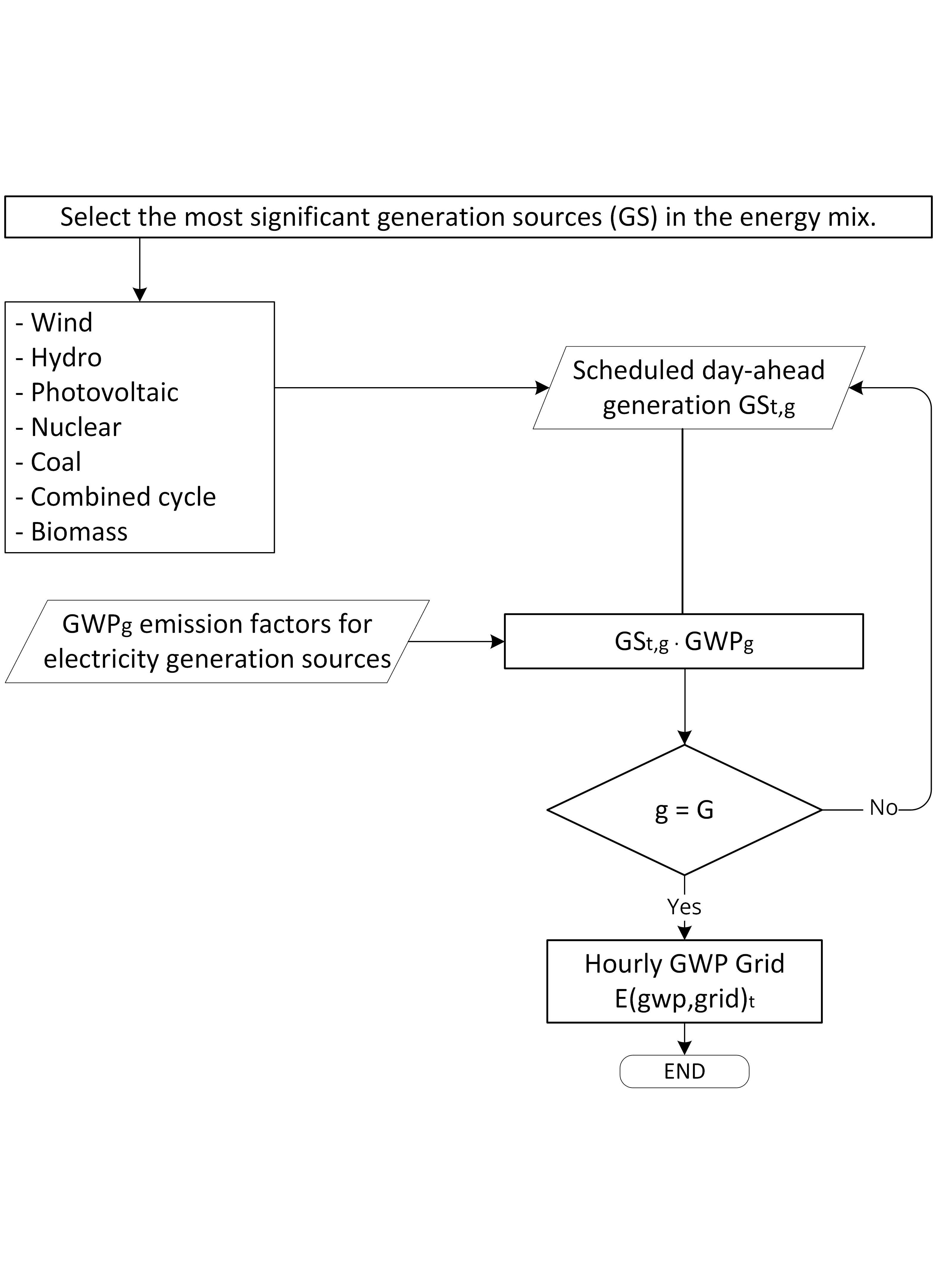}
  \caption{Methodology employed to calculate the hourly global warming potential of the grid.}
  \label{fig:ECmethodologygwp}
\end{figure}

For this article, the average GWP value is used. Figure \ref{fig:ECmethodologygwp} illustrates the methodology used to calculate the coefficients associated with the hourly emissions of the energy mix using values from Table \ref{tab:gwp_lec}. This process begins by selecting the generation sources with the highest poundage in the energy mix. Ideally, the generation sources should represent almost all the generated power so that the grid emissions indicators are as precise as possible to reality. Once the generation sources (GS) have been specified, the data for the next day's scheduled generation, which are public and available the day before their provisioning, are acquired. For each hour, the amount of scheduled generation for that technology $g$ is multiplied by its corresponding GWP emission factor. When calculated for the entire set of generation sources G, the total GWP of the network for the hourly time interval t is calculated. This process is repeated for each hour and day, successively.

\section{LEC optimization}

This section focuses on mathematically modeling the different assets that participate in the LEC, such as  photovoltaic generation, BESS, and the group of consumers who will benefit from this clean energy generation. First, the objective functions and system constraints are defined, followed by the presentation of the mathematical models for each component that takes part in the LEC.

\subsection{Objective functions}

\textbf{Price-based}

This program focuses exclusively on the economic aspect. It minimizes the overall LEC electricity bill (\ref{eq:objectivefunction_pricebased}), considering the centralized battery degradation cost $K_{t}^{cal}$ due to calendar aging, where $P_{t}^{buy}$ and $P_{t}^{sell}$ are the time-varying price for buying and selling electricity. The variables $\chi_{t,p}^{buy}$ and $\chi_{t,p}^{sell}$ refer to the virtual energy purchased and sold to the grid by each participant $p$, and $P^{VAT}$ is the tax used. The objective function is expressed as

\begin{equation}
    \text{min } {f_{1}=\sum_{t=1}^{T} \sum_{p=1}^{P} (P_{t,p}^{buy}\chi_{t,p}^{buy} - P_{t,p}^{sell}\chi_{t,p}^{sell} + K_{t}^{cal}}) \label{eq:objectivefunction_pricebased}
\end{equation}

\textbf{Environmental-based}

This optimization strategy attempts to minimize the carbon footprint occasioned by the generation sources that provide electricity to each participant involved in the LEC. In this approach, only the emissions from the energy consumed are considered, not those sold to the grid. The objective function is formulated as follows.

\begin{equation}
 \text{min }{f_{2}=\sum_{t=1}^{T} \sum_{p=1}^{P}(E_{t}^{gwp,grid} \chi_{t,p}^{buy}) + E^{gwp,pv} W_{t}^{pv} + E^{gwp,bat}\sigma_{t}^{dis}} \label{eq:EBobjective}
\end{equation}

where $E_{t}^{gwp,grid}$ indicates the kg $CO_{2-eq}$/kWh of the grid on average per period $t$. It is calculated with the hourly energy production mix, taking the values of scheduled generation in the day-ahead market for each technology described in Table \ref{tab:gwp_lec}. The emission parameters associated with the battery $E^{gwp,bat}$ and photovoltaic $E^{gwp,pv}$ are indicated also in Table \ref{tab:gwp_lec}.
\\

\subsection{LEC constraints}

\textbf{Energy balance}

The energy balance of the Local Energy Community allows for distinguishing the energy generated through the centralized PV and BESS and imported from the grid. In this case, a sharing coefficient $\beta_{t,p}$ is virtually associated with each participant. Therefore, if a participant $p$ has a constant/static sharing coefficient of $\beta_{t,p}=0.5$, it means that this consumer owns 50\% of the electricity generated by the renewable generation of the community. 

The energy balance expressed in \ref{eq:energybalance} is designed to simulate that the participants have a battery and photovoltaic generation. In order to achieve a balanced energy system, the total electricity imported from the grid 
 $\chi_{t,p}^{buy}$, must balance the production from generation units, consumption from load units, charging and discharging of the central BESS and energy sold for each period of time $t \in T$: 

\begin{equation}\label{eq:energybalance}
\beta_{t,p}(W_{t}^{pv}+\sigma_{t}^{dis}-\sigma_{t}^{ch}) + \chi_{t,p}^{buy} = W_{t,p}^{inflex,load} +  \chi_{t,p}^{sell}
\end{equation}


\textbf{Not buy and sell at the same time}

Binary variables $\delta_{t,p}^{buy}$ and $\delta_{t,p}^{sell}$ are now introduced in order to ensure that it is not possible to sell and buy in the same period for each participant. It is possible that when visualizing the total consumption of the LEC, there may be periods where consumption and selling occur simultaneously. However, this is due to the fact that one participant is selling while another is not. This occurs because each client may have different distribution coefficients, resulting in one having surpluses while the other does not.

\begin{equation}\label{eq:lec-individual-emport-export}
   \delta_{t,p}^{buy} + \delta_{t,p}^{sell} \leq 1
\end{equation}

\textbf{Prosumers capacity limits}

Electricity bought and sold must be below power limits, according to the terms stipulated in the retail contract: 

\begin{equation}\label{eq:lec-buy}
\chi_{t,p}^{buy} \leq \delta_{t,p}^{buy} \cdot X_{p}^{max,import} 
 \end{equation}

\begin{equation}\label{eq:lec-sell}
\chi_{t,p}^{sell} \leq \delta_{t,p}^{sell} \cdot X_{p}^{max,export}
\end{equation}

\textbf{Net generation}

The net-generation $\theta_{t}$ follows this equation
\begin{equation}
    \theta_{t}^{lec} = W_{t}^{pv} +  \sigma_{t}^{dis}  -\sigma_{t}^{ch}
\end{equation}

The individualized net hourly energy generated by those energy community participants $p$ that carry out collective self-consumption, $\theta_{t,p}$, is

\begin{equation}
    \theta_{t,p} = \beta_{t,p}\theta_{t}^{lec}
\end{equation}

where $\theta_{t}^{lec}$ represents the total hourly net energy produced by the generator smart meter and $\beta_{t,p}$denotes the hourly distribution coefficient among consumers participating in the collective self-consumption of the energy generated, in period $t$. When referring to the static sharing coefficient, the same value is repeated for all $t$. Consumers are required to submit the coefficients of participants involved in self-consumption for all hours of the current year, which cannot be modified within the same year.

\textbf{Sharing strategy}
The sum of all sharing coefficients allocated to the LEC participants for each time period $t$ must be equal to 1.

\begin{equation}
    \sum_{p=1}^{P} \beta_{t,p} = 1
\end{equation}

\section{Case study and results}

\begin{figure}[!]
  \centering
  \includegraphics[width=0.5\textwidth]{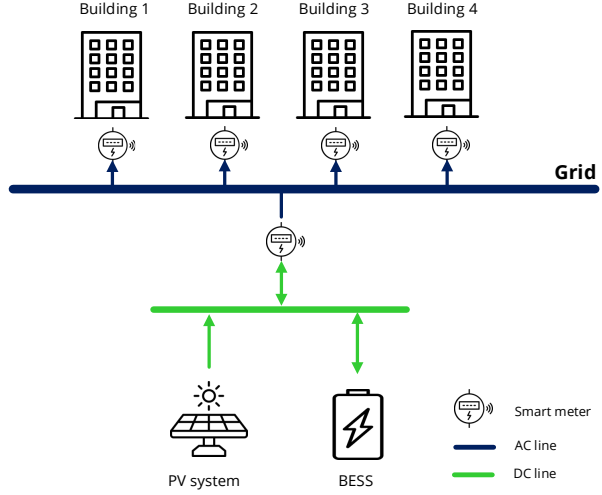}
  \caption{Case study local energy community simplified scheme.}
  \label{fig:BEMS_scheme}
\end{figure}

This study is focused on a renewable energy community located in Catalonia, consisting of four office buildings and a collective-owned centralized photovoltaic system and a Li-ion BESS. These components are connected downstream of an inverter that feeds energy back into the grid, which is then compensated in the electricity bill of each participant based on their assigned generation percentage. In addition, this smart meter is bidirectional and capable of purchasing electricity from the grid when it is necessary to store energy in the battery. Each building has its own smart meter and purchases energy from the grid, since they have not flexible assets behind the meter. Figure \ref{fig:BEMS_scheme} illustrates the components of the renewable energy community: four office buildings and a centralized PV and BESS as a collective renewable generation unit. The Local Energy Community Management System is responsible for controlling and optimizing the LEC flexible asset and must send optimal operating set-points for the centralized BESS, which is the only flexible source as illustrated in Figure \ref{fig:BEMS_scheme}. 

This case study is based on an actual energy community located in Spain. The data available for PV generation and building consumption ranges from March 3 to December 31, 2022. According to the European and Spanish regulations, this study belongs to the category of a renewable energy community with collective self-consumption and surplus compensation \cite{Disposic10:online}. In other words, the energy generated is used to supply multiple points of consumption. This group of participants agrees to distribute the renewable energy generated and applies static sharing coefficients for each consumer. For each end-user, any surplus energy that is not consumed is fed back into the grid and economically compensated on its electricity bill.

\subsection{Energy Community specifications}

The BESS parameters and specifications used in this study are listed in Table \ref{table:lec-soc}. The optimization imposes that the state of charge of the battery is the same at the beginning $\sigma_{t=0}^{soc}$ and at the end of the optimization horizon  $\sigma_{t=end}^{soc}$, which is 24 hours ahead in this study. This prevents the battery from being completely discharged at the end of the day. 

{
\renewcommand{\arraystretch}{1.5}
\begin{table}[htbp]
\centering
\footnotesize
\caption{LEC BESS parameters.}\label{table:lec-soc}
\begin{tabular}{ll}
\hline
\textbf{Input Parameters}         & \textbf{Value} \\ \hline
Maximum charging power allowed    & 90 kW                 \\
Maximum discharging power allowed & 90 kW                 \\
Maximum SOC                       & 189.9 kWh             \\
Minimum SOC                       & 31.65 kWh             \\
Efficiency charging               & 0.95                 \\
Efficiency discharging            & 0.95                 \\
$\sigma_{t=0}^{soc}$                        & 150 kWh               \\
 $\sigma_{t=end}^{soc}$                     & 150 kWh               \\ \hline
\end{tabular}
\end{table}
}

Table \ref{tab:building-params} lists the static parameters related to each building that constitutes the energy community and Table \ref{tab:building-shared-coeff} indicates the static sharing coefficient rate of generation corresponding to each building.

{
\renewcommand{\arraystretch}{1.5}
\begin{table}[htbp]
\centering
\footnotesize
\caption{LEC participants maximum contracted power.}\label{tab:building-params}
\begin{tabular}{lllll}
\hline
\textbf{Input Parameters}         & \textbf{B1} & \textbf{B2} & \textbf{B3} & \textbf{B4} \\ \hline
Spanish electricity tariff type   & 3.0                 & 3.0                 & 3.0                 & 3.0                 \\
Maximum contracted power Period 1 & 70                  & 43.65               & 20.785              & 75                  \\
Maximum contracted power Period 2 & 70                  & 43.65               & 20.785              & 75                  \\
Maximum contracted power Period 3 & 70                  & 43.65               & 20.785              & 75                  \\
Maximum contracted power Period 4 & 70                  & 43.65               & 20.785              & 75                  \\
Maximum contracted power Period 5 & 70                  & 43.65               & 20.785              & 75                  \\
Maximum contracted power Period 6 & 70                  & 43.65               & 20.785              & 75                  \\\hline
\end{tabular}
\end{table}
}

{
\renewcommand{\arraystretch}{1.5}
\begin{table}[htbp]
\centering
\footnotesize
\caption{LEC participants static sharing coefficient.}\label{tab:building-shared-coeff}
\begin{tabular}{lllll}
\hline
\textbf{}         & \textbf{B1} & \textbf{B2} & \textbf{B3} & \textbf{B4} \\ \hline
Static sharing coefficient   & 0.35                 & 0.15                 & 0.02             & 0.48          \\
\hline
\end{tabular}
\end{table}}


\subsection{Grid GWP calculation}

\begin{figure}[h]
  \centering
  \includegraphics[width=1\textwidth]{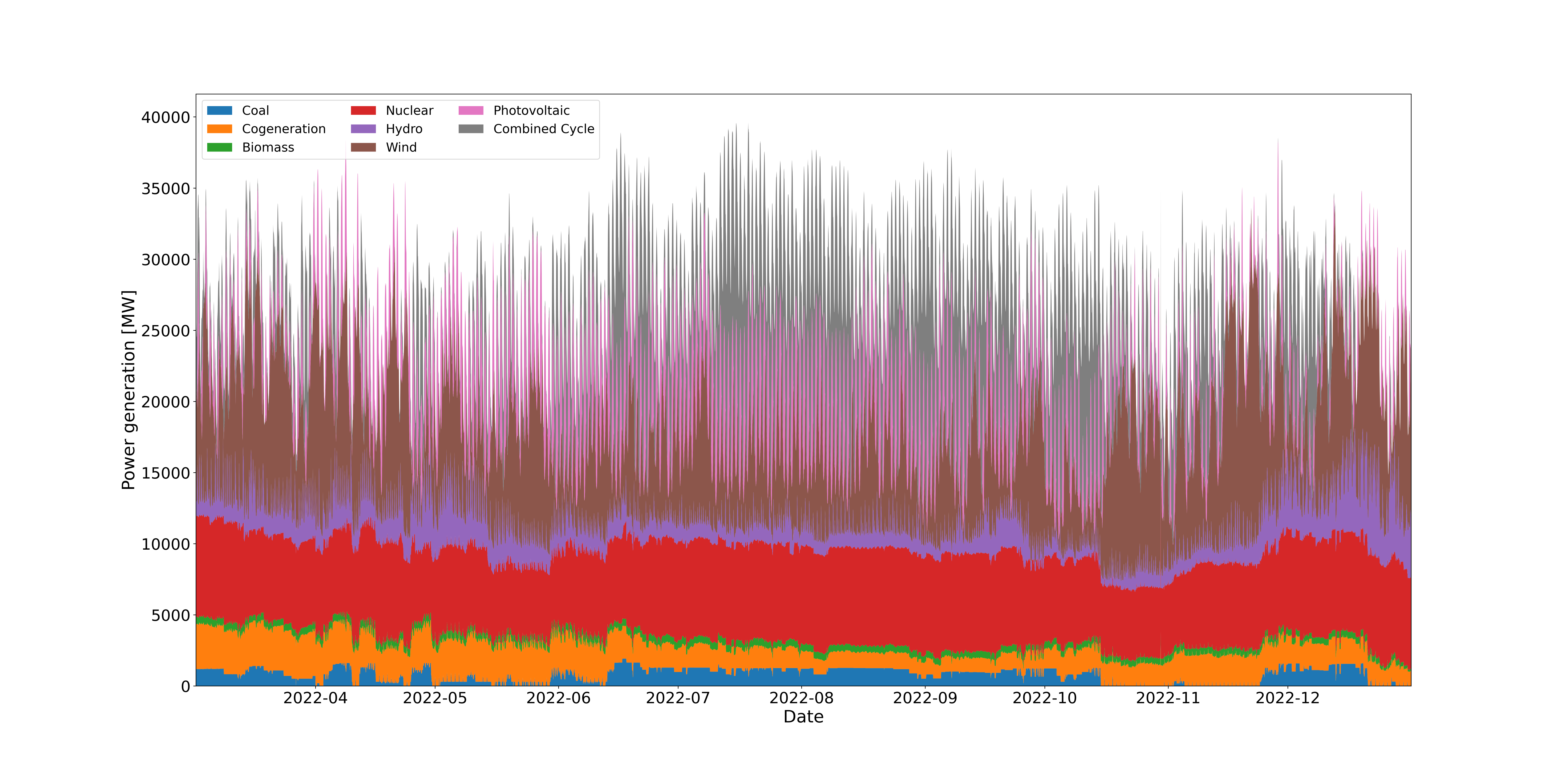}
  \caption{Generation sources energy mix sharing for the case study.}
  \label{fig:CH4-generation-share}
\end{figure}

The proposed case study and scenarios used data from March 3rd - December 31st, 2022, the available date range from the energy community. The hourly share of each generation source (see Table \ref{tab:gwp_lec}) in the Spanish energy mix is illustrated in Figure \ref{fig:CH4-generation-share}. The selected generation types are primarily responsible for the total generation and cover 96\% of the total. The presence of coal is almost nonexistent, but during periods of high energy demand, the energy produced from it increases. Nuclear power serves as the base generation. In summer, photovoltaic and combined cycle generation significantly increase, mainly due to the rise in demand caused by high temperatures in most regions of the country. Hydroelectric generation increases its share in the spring months and late winter months.

\begin{figure}[!]
  \centering
  \includegraphics[width=0.8\textwidth]{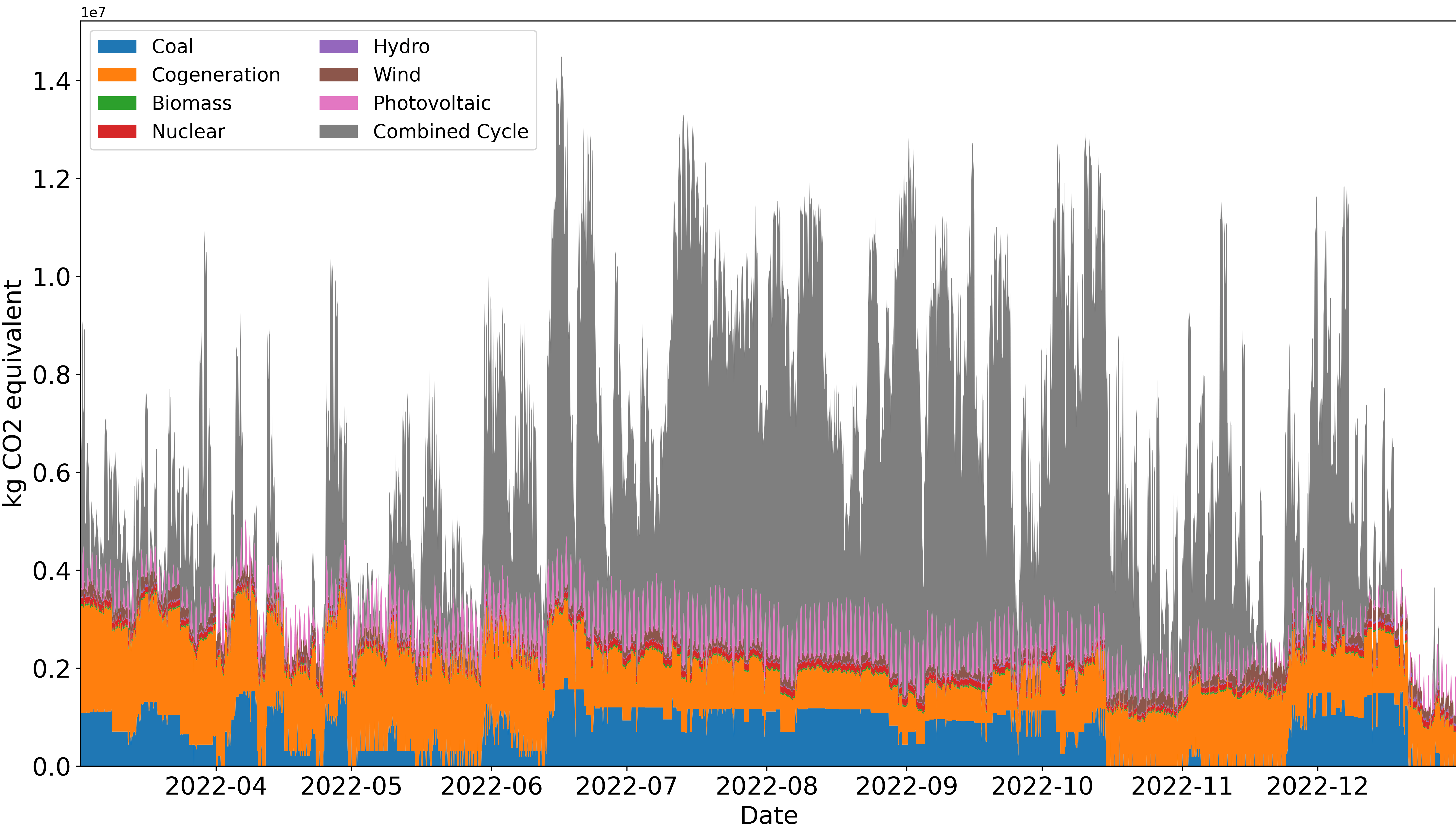}
  \caption{Spanish energy mix kg $CO_{2}$ equivalent in the energy mix.}
  \label{fig:grid-gwp}
\end{figure}

The hourly grid GWP is calculated and shown in Figure \ref{fig:grid-gwp}. As expected, fossil-based generation sources such as coal, combined cycle and cogeneration contribute the most to the greenhouse gas emissions in the energy mix, despite not being the generation sources with higher production. For instance, although the energy produced from combined cycle cogeneration and coal in the mix is not the most significant, they are the sources that contribute the most $CO_{2}$ emissions to the grid.

\begin{figure}[h]
  \centering
  \includegraphics[width=0.9\textwidth]{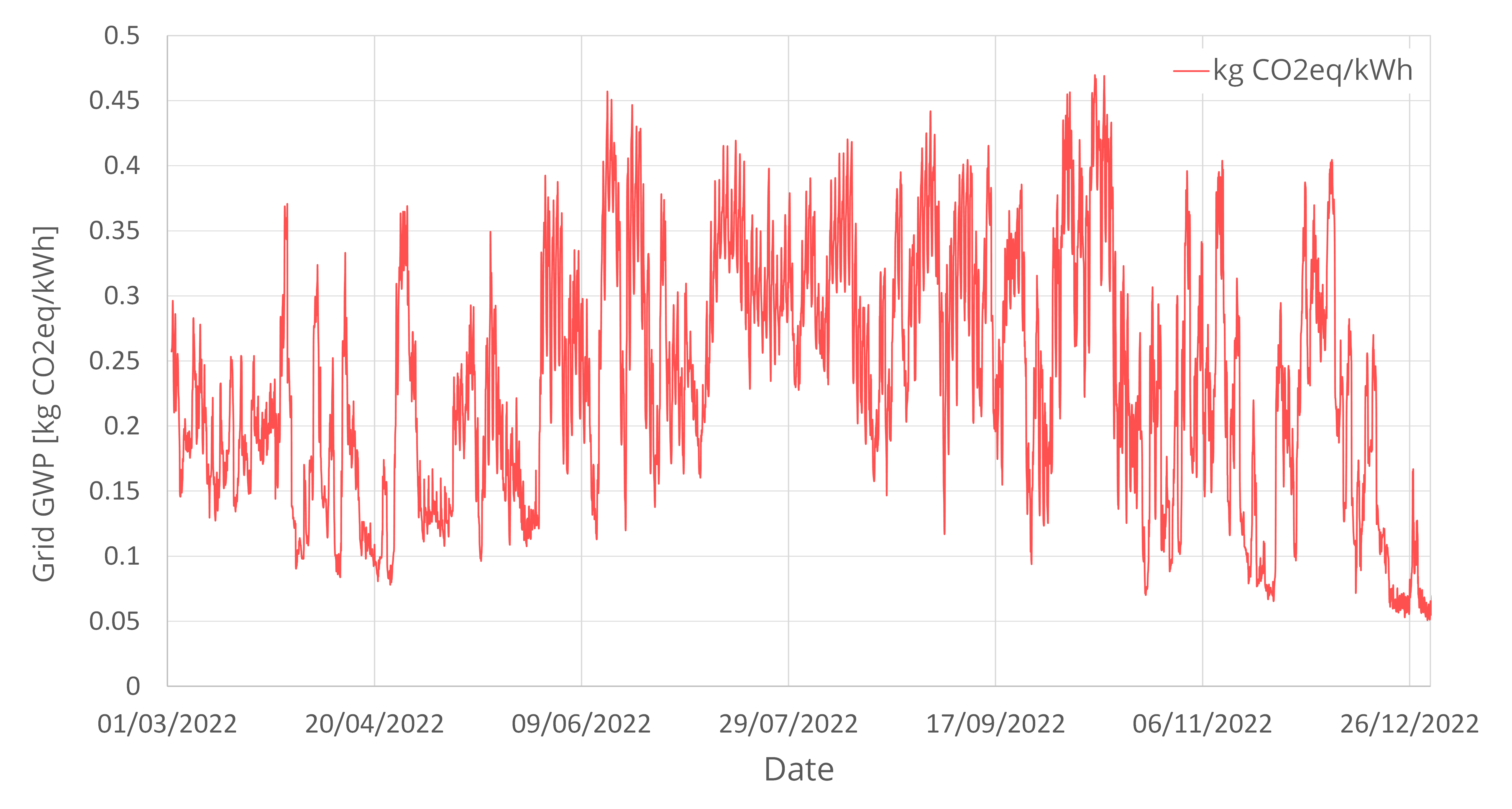}
  \caption{Spanish GHG emissions in the energy mix grid.}
  \label{fig:gwp-grid-energy-mix}
\end{figure}

Finally, the hourly grid GWP is calculated and displayed in Figure \ref{fig:gwp-grid-energy-mix}. Summer months display higher emissions associated with the grid energy mix, mainly due to the increased demand during these months and lower production of specific renewable resources such as hydro and wind power. However, thanks to the increased presence of photovoltaic generation in the last years, emissions have been partially reduced during the summer months in comparison to other past years. Low levels of emissions are usually associated with low energy demand in the system (i.e., mild temperatures) and high penetration of renewables, especially wind power. 

\section{Results}


This section presents the results of the different scenarios analyzed in the proposed case study. To enhance the clarity and comprehensibility of the results, only two representative days are displayed to illustrate the behavior of the centralized battery and energy community for the price and enviroment optimization strategies. Additionally, this section includes an analysis of the energy community costs, as well as the GHG emissions associated with their consumption for the case study period. This information is valid in order to evaluate the effectiveness of the proposed optimization strategies.

\subsection{LEC baseline consumption}

Prior to presenting the results of the two optimization strategies for the LEC, the cost and GHG emissions associated with the consumption of its participants are shown, assuming they did not have access to collective generation and had to purchase all energy directly from the grid. The aim is to enable a comparison of the effectiveness of the optimization developed based on these baseline data. Table \ref{tab:lec-baseline} shows the cost and GHG emissions costs and GHG emissions associated with each building over the 10-month study period. Thus, the total LEC electricity cost is 50 136.45 \EUR{}, and the emissions associated are 61.94 tones of $CO_{2-eq}$.

\begin{table}[]
\centering
    \caption{Baseline LEC cost and GHG emissions.}
\begin{tabular}{lrr}
\toprule
\textbf{Building} & \textbf{Total cost} & \textbf{Total GHG emissions} \\
 & (\EUR{}) & (t $CO_{2}$) \\
\midrule
B1 &  13 414.72 & 18.81 \\
B2 & 6 004.96 & 7.04 \\
B3 & 390.46 & 0.47 \\
B4 & 30 326.32 & 35.63 \\
\midrule
\textbf{LEC} & \textbf{50 136.45} &\textbf{ 61.94} \\
\bottomrule
\end{tabular}
    \label{tab:lec-baseline}
\end{table}

\subsection{Price-based approach}

The price-based scenario aims to minimize the overall energy bill cost for participants of the LEC by utilizing the flexibility of the central BESS and the inflexible PV generation.

\begin{figure}[htbp]
  \centering
  \includegraphics[width=0.8\textwidth]{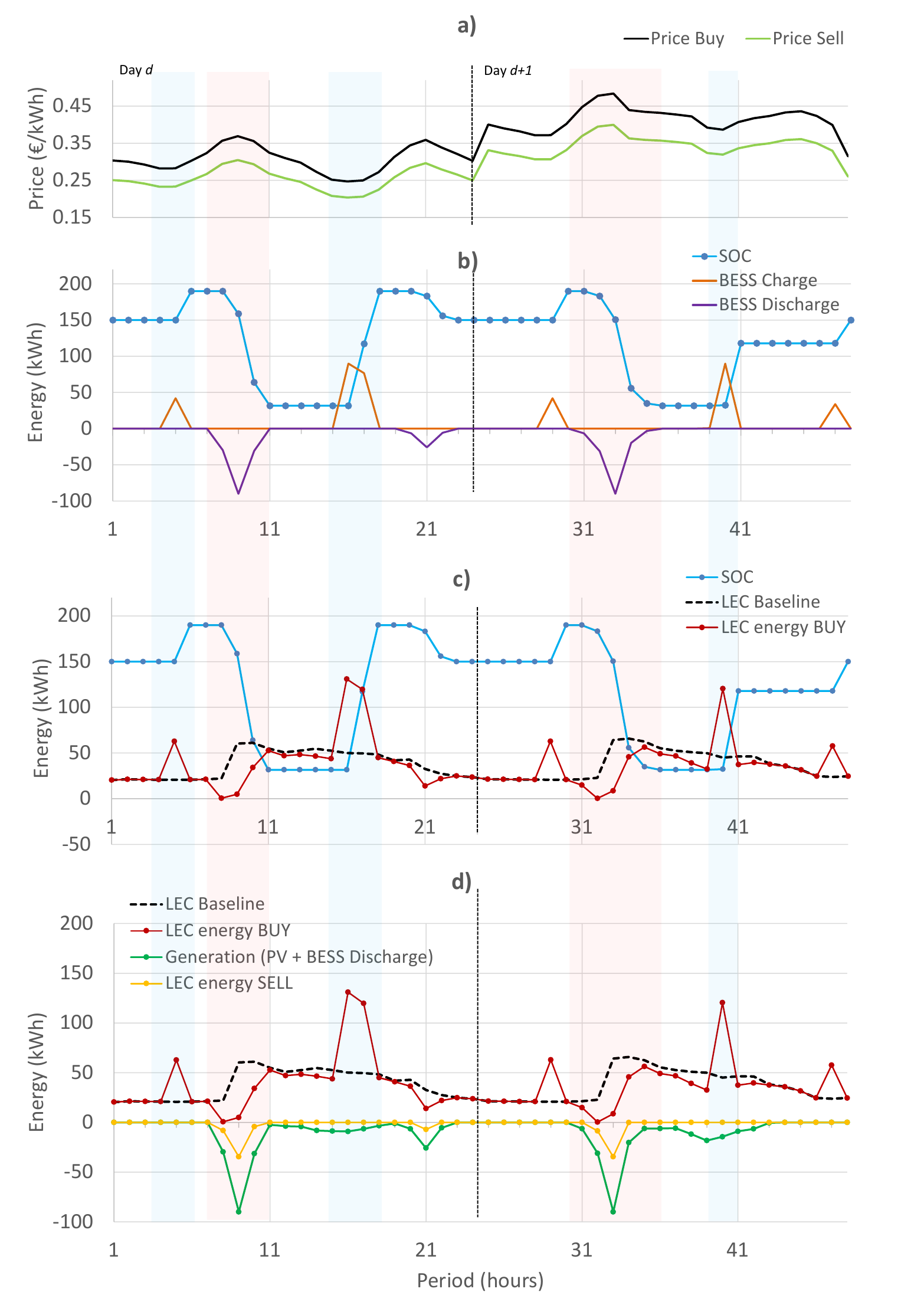}
  \caption{Price-based LEC optimization results. a) Electricity price for buying and selling to the grid. b) BESS performance c) LEC behavior after optimization d) LEC consumption, net generation produced and electricity sold.}
  \label{fig:ch4-pb-results}
\end{figure}

To demonstrate the feasibility and functionality of the developed Local Energy Management System (LEMS), its performance over two consecutive days is presented in Figure \ref{fig:ch4-pb-results}. This figure consists of four sub-figures, which are described and explained in detail through this sub-section.

The upper graph (Figure \ref{fig:ch4-pb-results} a)) displays the hourly price for buying and selling electricity on the grid. The purchase (Price buy) and self-consumption (Price sell) SPOT prices are used in this study. The buying price is always higher than the selling price because the taxes for purchasing are considered. Also, in this Figure, the start of the second day (Day \textit{d+1}) is indicated by a dashed black vertical line on all graphs to aid the reader. Additionally, to facilitate comprehension, periods with low electricity prices from the grid are highlighted in blue, while periods with higher prices are emphasized in light red. These elements will help with the visual explanation.

The following image (Figure \ref{fig:ch4-pb-results} b)) displays the behavior of the centralized battery of the LEC. It can be observed that during periods with high purchase prices from the grid, the battery discharges to reduce the costs associated with buying energy from the grid for the users of the community and/or selling excess energy to obtain economic remuneration in return.

Figure \ref{fig:ch4-pb-results} c) shows the state of charge of the centralized battery, the baseline consumption of the energy community (i.e., the actual and inflexible consumption of all buildings that are part of the LEC), and the consumption associated with the LEC, considering centralized generation sources (battery and photovoltaic generation). During periods of low prices (blue time slots), the LEC takes advantage of buying energy from the grid to charge the battery and use this energy later during periods of high prices. Therefore, those consumption peaks are the battery charge, but they are associated with each user, as the cost is distributed among them, using the corresponding distribution coefficient. On the other hand, during periods of high prices (red time slots), the battery discharges to alleviate the cost for the participants of the LEC. Thus, the consumption of the LEC is below the baseline, thanks to photovoltaic and battery discharge.

Finally, Figure \ref{fig:ch4-pb-results} d) shows the generation sources (battery discharge and PV) and the energy sold to the grid, for which participants will receive economic benefits for the supplied energy. To facilitate understanding of this graph, the author has chosen to assign negative values to the generation sources or energy sold and positive values to consumption sources. According to the amount of energy sold to the grid (see periods 9 to 11 on day \textit{d}), the battery discharges more than needed to meet user demand and sells this excess energy at a higher selling price. This occurs again during periods 21 (day\textit{ d}) and 32-34 (day \textit{d+1}). The reason is that selling energy to the grid is very profitable, making it worthwhile.

Overall, the case study results demonstrate the effectiveness of the LEMS in reducing costs and generating economic benefits for the energy community participants. The price-based results for 10 months of time horizon, and assuming perfect predictions, are summarized in the following Table \ref{tab:ch4-results-pb}. The objective is to reduce the overall LEC cost above the individual benefit. The LEC reduces the energy bill a 22.9\% (11 463.31 \EUR{}) in exchange for raising the GHG-associated emissions by 20.4\% (an increment of 12.64 tones of GHG emissions), mainly due to the BESS usage, since when it buys energy from the grid, the grid and BESS emissions are considered. 

\begin{table}[!]
\centering
    \caption{Price-based LEC cost and GHG emissions.}
\footnotesize
\begin{tabular}{lrrrr}
\toprule
\textbf{Building} &  \textbf{PB Total cost} &   \textbf{Baseline} & \textbf{PB Total GHG} & \textbf{Baseline}  \\
 & (\EUR{}) && (t $CO_{2}$)& \\\midrule
B1 &  9 529.01 & $\downarrow$ 29.0\% &  21.65 &  $\uparrow$ 15.1\% \\
B2 & 4 301.87 & $\downarrow$ 28.4\% & 11.15&  $\uparrow$ 58.3\%\\
B3 & 184.17 & $\downarrow$ 52.8\% & 5.64 &  $\uparrow$ 1108.2\%\\
B4 & 24 658.09 &$\downarrow$ 18.7\% & 36.14 &  $\uparrow$ 1.4\%\\
\midrule
\textbf{LEC} & \textbf{38 673.14} & $\downarrow$  22.9\% &\textbf{74.58} & $\uparrow$  20.4\%\\
\bottomrule
\end{tabular}
    \label{tab:ch4-results-pb}
\end{table}

It should be mentioned that the installed photovoltaic generation covers, on average, 17\% of the total LEC consumption; thus, the savings derived from photovoltaics are not very noticeable. However, thanks to the battery, which purchases energy during low-cost periods and injects it into the grid during high-cost periods, significant savings can be achieved.


\subsection{Environment-based approach}

This optimization strategy aims to minimize the GHG emissions associated with the electricity consumption of the energy community participants. To quantify these emissions, the GHG of the grid is considered, which varies hourly depending on the energy mix, centralized PV generation, and BESS use.

To minimize greenhouse gas emissions associated with consumption, two different days are selected to explain the different behavior of the LEMS depending on the energy mix: the first  with a high penetration of renewables in the grid, resulting in a low GWP index, and the second with a high percentage of fossil sources in the energy mix.

\subsubsection{Low grid GWP}

\begin{figure}[!]
  \centering
  \includegraphics[width=0.8\textwidth]{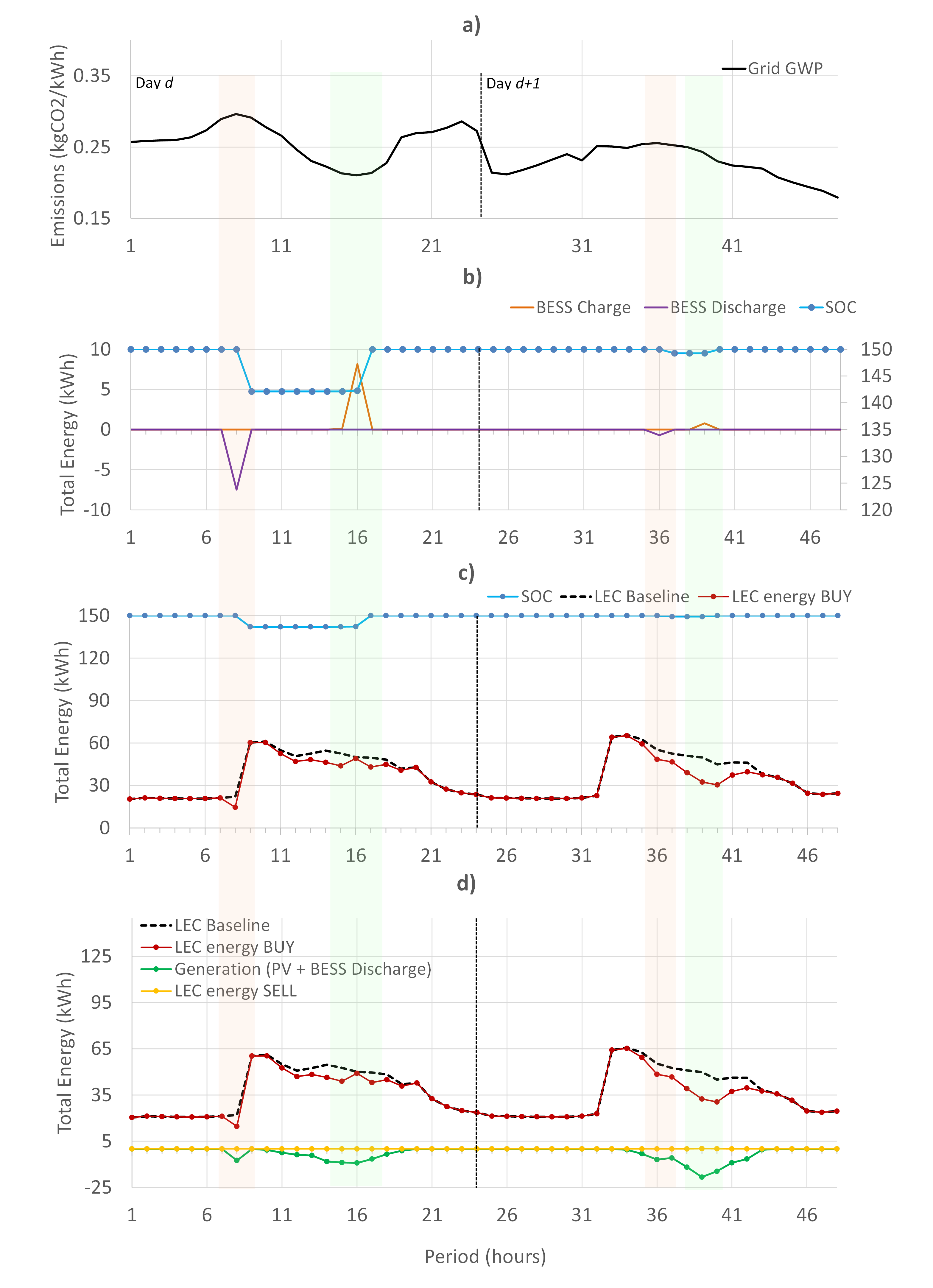}
  \caption{Methodology}
  \label{fig:ch4-eb-results-low}
\end{figure}

Following the same format as the previous approach, Figure  \ref{fig:ch4-eb-results-low} presents the results of the energy community for two days of optimization.

The figure above (Figure  \ref{fig:ch4-eb-results-low}a) shows the emissions curve associated with the hourly energy mix of the grid. The higher the curve, the greater the emissions associated with consumption if energy is purchased from the grid.

Figure \ref{fig:ch4-eb-results-low} (a) above shows the emissions curve associated with the hourly energy mix of the grid. The higher the curve, the greater the emissions associated with consumption if energy is purchased from the grid.

In the second image, Figure  \ref{fig:ch4-eb-results-low}(b), the battery is discharged during periods of the day with higher pollution indices (period 8), thus reducing emissions associated with consumption. As the battery capacity must be the same at the beginning and end of the day (150 kWh), surplus photovoltaic energy is used to charge the battery during period 16. On day \textit{d+1}, the battery is slightly discharged during the period of maximum GHG emissions in the grid and charged with surplus solar energy. On day \textit{d+1}, the battery is hardly used because, thanks to the high penetration of renewables, it is cleaner to purchase energy from the grid.

In Figure  \ref{fig:ch4-eb-results-low}(d), it can be observed that photovoltaic energy supplies part of the local energy community's consumption, and in specific periods, this photovoltaic energy is used to charge the battery.

Overall, the case study demonstrates the potential benefits of combining renewable energy sources with battery storage systems to reduce greenhouse gas emissions and increase energy efficiency.

\subsubsection{High grid GWP}

\begin{figure}[htbp]
  \centering
  \includegraphics[width=0.8\textwidth]{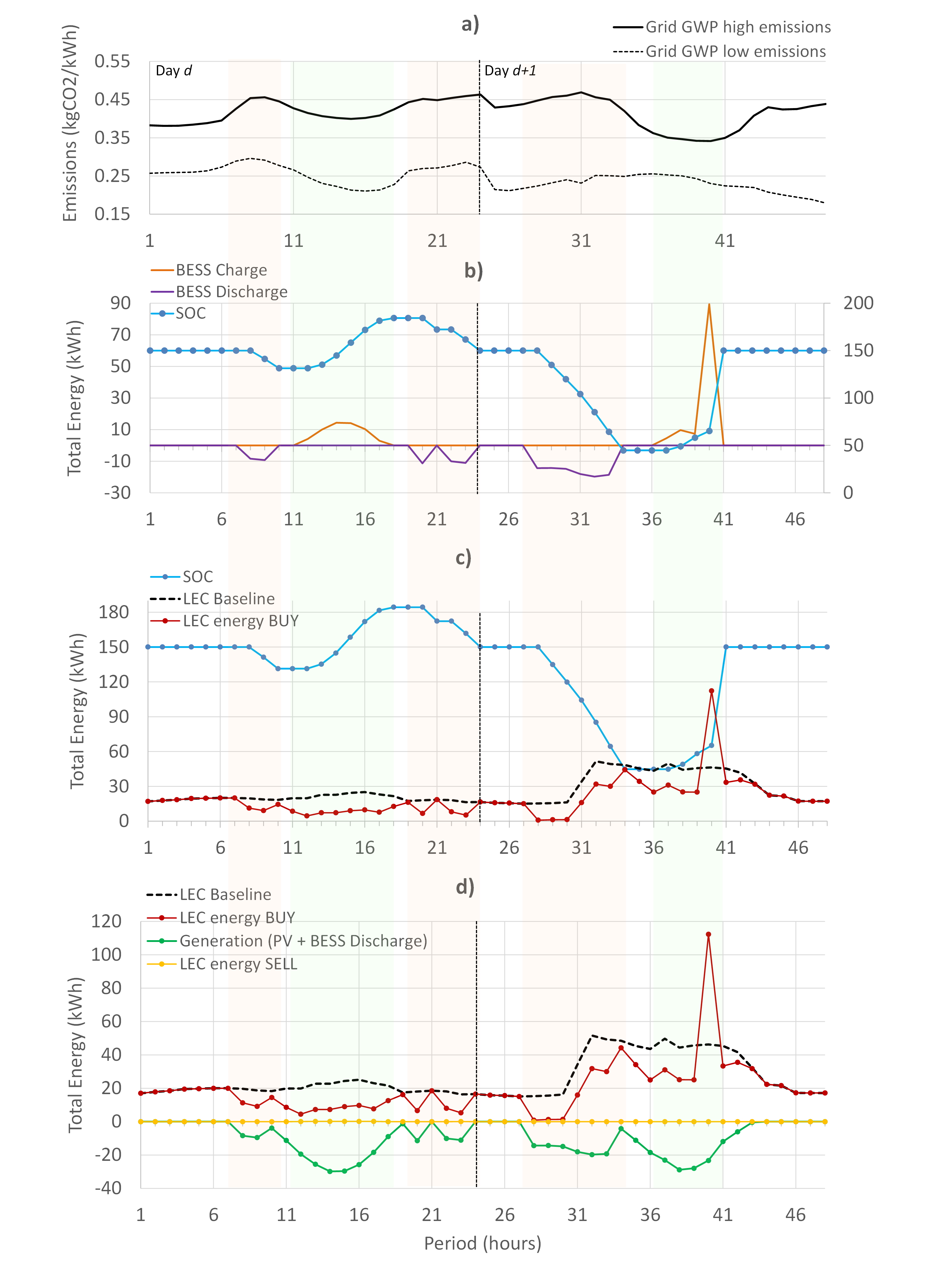}
  \caption{Methodology}
  \label{fig:ch4-eb-results-high}
\end{figure}

For this subsection, two days with a high percentage of fossil sources in the energy mix have been selected. Figure  \ref{fig:ch4-eb-results-high} a) shows the GWP with high GHG indices and the GWP of the previous example to compare both magnitudes.

At first glance, the main difference is that the use of the battery increases as the GWP index in the national energy system becomes higher. For the first day (d), the battery discharges during the period with the highest pollution associated with the grid. In the following periods, the battery recovers its state of charge thanks to photovoltaic generation. From period 21-24, the battery discharges, coinciding with the hours of highest emissions on the grid. For the next day (day d+1), the battery starts with the initial state of charge of 150 kWh, as the restriction imposes. The BESS discharges from 28 to 34, also coinciding with a period of high emissions. From there, the battery recovers its SOC thanks to the photovoltaic energy from the LEC and by charging it in the period 42 at maximum power (90 kW) to meet the SOC restriction. This charge is made during the period where emissions are the lowest during that day.

In Figure \ref{fig:ch4-eb-results-high} c), it can be observed that the LEC only makes a greater electricity purchase than the baseline to charge the battery in period 40. As mentioned before, this period for day d+1 is the one with the lowest CO2.

In Figure \ref{fig:ch4-eb-results-high} d), it can be seen that no energy is sold to the grid and that the photovoltaic generation covers part of the participants' demand. During the night and in periods of high emissions, the battery discharges part of its capacity to reduce the emissions associated with consumption.

\begin{table}[!]
\centering
    \caption{Environment-based LEC cost and GHG emissions.}
\footnotesize
\begin{tabular}{lrrrr}
\toprule
\textbf{Building} &  \textbf{EB Total cost} &   \textbf{Baseline} & \textbf{EB Total GHG} & \textbf{Baseline}  \\
 & (\EUR{}) && (t $CO_{2}$)& \\\midrule
B1 &  10 131.23 & $\downarrow$ 24.5\% &  16.82 & $\downarrow$ 10.6\% \\
B2 & 4 590.18 & $\downarrow$ 23.6\% & 7.30& $\uparrow$ 3.7\%\\
B3 & 210.17 & $\downarrow$ 46.2\% & 2.28 & $\uparrow$ 3.87\%\\
B4 & 25 757.76 & $\downarrow$ 15.1\% & 31.88 & $\downarrow$ 10.5\%\\
\midrule
\textbf{LEC} & \textbf{40 689.35} & \textbf{$\downarrow$ 18.8\% }&\textbf{58.27} & \textbf{$\downarrow$ 5.9\%}\\
\bottomrule
\end{tabular}
    \label{tab:ch4-results-pb}
\end{table}

\section{Conclusions}

The LEC optimization results demonstrate the feasibility and satisfactory performance of the approaches proposed in this thesis. On the one hand, the price-based optimization achieves savings of approximately 23\% (11,500 \EUR{}) compared to the LEC baseline consumption. The contribution of photovoltaics compared to consumption is lower than this percentage, so the battery plays a crucial role in achieving these savings. However, the associated emissions increase by 20\%. On the other hand, the strategy of minimizing emissions associated with consumption achieves a reduction of approximately 6\% of these contaminants compared to the baseline consumption, avoiding the emissions of 3.7 tons of $CO_{2-eq}$. The difference in GHG release between the price and environmental-based strategies is 21\%.

The positive aspect is that the environmental strategy, in addition to minimizing emissions, is capable of reducing costs by 18\% (approximately 10,000 \EUR{}) compared to the baseline, making it feasible to choose to enhance the environment without incurring high costs, which is a significant advantage. However, there are still some challenges confronting the proliferation of LECs. A significant challenge is regulatory barriers due to the complex framework required to obtain the necessary permits and approvals for installation and operation. This can make it difficult for LECs to secure financing and access to the energy grid, as conventional investors may be uncertain about investing in short-time tested business models.
Additionally, technical challenges associated with integrating renewable energy sources into the energy grid and storage systems are common. Also, at this early stage, LECs may lack the technical expertise to design, build, and operate renewable energy projects. Finally, due to a lack of awareness, many citizens may be unaware of the benefits and opportunities offered by energy communities, making it challenging to attract members and build critical support.

Despite these obstacles, LECs offer multiple benefits to neighborhoods, the environment, and provide promising investment opportunities. One main advantage is the reduction of dependence on fossil fuels, resulting in a decrease in greenhouse gas emissions. By generating and consuming energy locally, LECs avoid losses due to the Joule effect and costly long-distance energy transmission and distribution lines. Furthermore, LECs can enhance energy security by reducing the vulnerability of areas to power outages and disruptions. They also promote community engagement and empowerment by enabling end-users and small groups to participate in the energy system, offering communities the possibility of generating income and having a more significant influence on the electricity market.

\section*{Acknowledgements}
This publication is part of the project MERIDIAN TED2021-131753B-I00, funded by MCIN/AEl/10.13039/501100011033 and by the European Union "NextGenerationEU/PRTR".
The work of Andreas Sumper was supported by the Catalan Institution for Research and Advanced Studies (ICREA) Academia Program.

\bibliographystyle{plainnat}
\bibliography{main}

\end{document}